\def\beq{\begin{eqnarray}}
\def\eeq{\end{eqnarray}}
\def\bea{\begin{eqnarray*}}
\def\eea{\end{eqnarray*}}

\documentclass{ws-procs9x6}

\begin{document}

\title{Supersymmetry, Naturalness
and the Landscape\footnote{\uppercase{T}his work supported in part
by the U.S. Department of Energy.}}

\author{M. Dine}

\address{Physics Department and Santa Cruz Institute for
Particle Physics \\
University of California \\ 
Santa Cruz, CA  95064\\ 
E-mail: dine@scipp.ucsc.edu}

\maketitle

\abstracts{
We argue that the study of the statistics of the landscape
of string vacua provides the first potentially predictive -- and
also falsifiable -- framework for string theory.  The question
of whether the theory does or does not predict low energy supersymmetry
breaking may well be the most accessible to analysis.  We argue
that low energy -- possibly very low energy -- supersymmetry
breaking is likely to emerge, and enumerate questions which must
be answered in order to make a definitive prediction.}

\section{Introduction:
Beyond the Standard Model in the late 20th Century}

In the last few years, evidenced has accumulated that there are a vast number
of metastable solutions of string theory.  This collection of states
has been dubbed the ``landscape" of string vacua.\cite{Susskind:2003kw}  This
development has, for understandable reasons, lead to
a great deal of criticism -- perhaps complaining would
be a more accurate description.
Complaining about this landscape is not a particularly productive
activity; it either exists as part of the underlying theory of quantum
gravity, or it does not.  If it does, its implications must be confronted.
I have been a vocal
skeptic\cite{Banks:2000pj,Banks:2003es} about the
existence of the landscape, and still have doubts, but
the evidence is impressive, and the possibility
demands serious attention.

Complaining is one of the great
pleasures of life, though, so let's examine the complaints.
Those who scoff at the landscape complain principally
that the landscape lacks predictivity.
I will try to argue that, at least compared with string theory as we have
practiced it for the past twenty years, {\it the landscape is the first
predictive framework we have encountered.}  It may be far less than
we hoped for\footnote{Some leading physicists have forcefully
argued to me that the landscape, on the one hand, is quite possibly
the correct picture of the fundamental theory of nature, and on the
other, quite disappointing.}, but
it may answer some of the questions
which particle physicists have struggled with
for almost three decades.   To this end, I would put forward the following
model of what physicists, thinking about what may lie beyond the Standard
Model, did for the quarter century after
the Standard Model was established.  I would divide us into two classes:
\begin{enumerate}
\item
 Model Builders:  Model builders explored an infinitely large space of possible 
field theories. (A {\it very} large infinity if allow extra dimensions, 
non-renormalizability...)  In this exploration, there were a few guideposts
and rules: phenomenological constraints, {\it naturalness, 
simplicity.}
\item  String Theorists:  String theorists confronted the
existence of a huge number of classical solutions.  Most of these
bear no resemblance to the world around us.  At the quantum level,
it is hard to exhibit solutions with $N<2$ supersymmetry.  Simple,
general arguments
showed that 
it would be difficult to find stable, non-susy (broken susy) vacua 
in any controlled approximation. There was also no (persuasive) clue to 
understanding the small value of the cosmological constant.  For a theory
which by its nature is a theory of gravity, this proved an enormous
obstacle.

In response to this situation, many string theorists simply dropped
any attempt to think about string phenomenology.  Those who persisted
were forced to adopt one of two approaches:  

\noindent a.  Look for realistic models at weak coupling.  Assume 
these are selected, and that the features
they exhibit at weak coupling survive at strong coupling (or that the couplings
are
accidentally weak).

\noindent b.  Look at generic features of string models (susy, 
axions, large dimensions), and hope these are somehow general, reflecting 
properties of some stable quantum system(s).
\end{enumerate}

Understand that I am not criticizing those working in these frameworks; I
have been a practitioner of all of these approaches.  But in thinking about the landscape,
it is helpful to keep in mind what we actually do.  One lesson, well known
to model builders, should be kept in mind:  almost all models {\it are wrong}.
We will see that the landscape can be a guide to model building -- one in
which notions of naturalness, simplicity and the like are sharp.  From the
perspective of a model builder, it is a new game which is fun; at worst,
it is, like most approaches, wrong.

From a string theory perspective, I will argue that for the first time the landscape
provides a predictive set of rules.  There are a few questions which we will
not attempt to answer:  the value of the cosmological constant, probably the value
of the weak scale and certain couplings; we will have to write these quantities
off as ``anthropic" or ``environmental."  But the fact that we can
even accommodate the
observed values of these quantities is something new and exciting.  Some
might argue that {\it  the most
disappointing aspect of the anthropic solution of these problems is perhaps
that it is so pedestrian.}  But this is the first time in string theory we have
had any way of understanding these seemingly fundamental questions.  Perhaps there
is a deeper, more beautiful answer, but, lacking that,
it doesn't make sense to ignore one which
we have.

\section{An Overview of The Flux Landscape}

The idea of a discretuum of states, which might
provide an understanding of the small value of the cosmological
constant,
has a rather long history.\cite{Weinberg:1988cp,Banks:1991mb,Bousso:2000xa,Feng:2000if}
But until the work of Kachru, Kallosh, Linde
and Trived(KKLT),\cite{Kachru:2003aw} while there were a variety of scenarios, there were no
persuasive constructions in anything resembling a systematic
approximation.  KKLT exhibited metastable points in the moduli effective 
potential, in controlled (or nearly controlled) approximations. These
states were both dS and AdS, with and without supersymmetry.
Their work strongly 
indicated the existence of a vast number with all moduli stabilized. 
There are many questions one can raise about this analysis.
Most importantly, all of these states must be understood cosmologically.
Somewhere in their past or future they have various singularities,
and we don't know how or whether these might be resolved.
Moreover, the notion of connectedness in this landscape is, at best,
obscure.\cite{Banks:2003es}  But it is not clear that these obstacles are
insurmountable,\cite{Freivogel:2004rd} and few physicists -- even
those who find the landscape repugnant -- seem to take these objections
seriously.  This is, in part, because they apply to virtually any
string theory configuration with less than four supersymmetries.
Still, the possibility that the landscape may not
exist should be kept in mind.

Accepting, at least provisionally, the existence of the landscape, the nature and goals of
string theory (fundamental physics) are different than we previously
imagined.  
In this vast ``landscape", one can't hope to find ``the state" 
which describes our universe.  Our interest shifts to
the statistics of these 
states.\cite{Douglas:2003um,
Ashok:2003gk,Douglas:2004kp}  Nor can our goal to be predict all of the
features of nature with arbitrary precision.  But there is a very
real possibility of predictions, based on finding {\it correlations}
among properties of the states.  One also has the possibility
for falsification:  typical states in the landscape might
be inconsistent with experiment.

We don't yet all we need to know to make predictions.  But
what is particularly remarkable is
that, thanks to the work of a number
of researchers, some features of the statistics
of the landscape are starting to emerge.
We can already do some prototype calculations, and can 
pose sharp questions, which can plausibly be answered.

One question looks particularly important and quite possibly
accessible:  does 
this framework predict low energy supersymmetry?  If so, does it 
suggest a particular scale for the breaking?  I will focus
on this possible prediction for future experiments,
as well as on one possibility for falsification:
   \begin{itemize}
   \item
   The possibility that cosmological constant + the hierarchically
   small value of the weak scale imply
   low energy supersymmetry.
   \item
   The problem that $\theta_{qcd}$ seems, within
   the landscape, to be a uniformly distributed random
   variable.\cite{Banks:2003es,Donoghue:2003vs}
   \end{itemize}

\section{Review of the KKLT Construction}

In this section, we briefly review the KKLT
construction.  We focus on a particular case:  orientifolds of IIB theory on a 
Calabi-Yau space.  In such theories, there are a variety of moduli:
complex structure moduli ($z_a$), Kahler moduli ($\rho_i$), 
and the axion-dilaton multiplet:  $~~~~~~\tau = {1 \over g_s} + ia$

IIB theory has RR and NS-NS three-index antisymmetric tensor 
fields, $F,H$. Solutions of the string equations exist on CY spaces 
with non-trivial, quantized fluxes, characterized by integers:
$$ \int_{\Sigma_i} H= M_i ~~~~~\int_{\Sigma_i} F = K_i$$
Here the integrals are over three-cycles, $\Sigma_i$.
In general, there are many (100's) of 
possible cycles.  There are also many possible values of the 
integers $K$ and $M$.  For generic fluxes, the
$z_a$'s and $\tau$ are fixed in these solutions.  
This has a low energy explanation:  
in the presence of flux, there is a non-trivial superpotential:\cite{Gukov:1999ya}
$W(z,\tau)$, at the leading order in the 
$\alpha^\prime$ (large radius) expansion.

An interesting example, which illustrates some
features which will be important in our later discussion
was provided by Giddings, Kachru and Polchinski (GKP),\cite{Giddings:2001yu}.
They studied a Calabi-Yau space near a conifold point in the
moduli space, focussing on the modulus, $z$, which
measures the distance from the conifold point.
With fluxes on collapsing three cycles at this point,
one finds both stabilization
and warping.  The superpotential is given by:
\beq W= (2 \pi)^3 \alpha^\prime (M \mathcal{ G}(z) -K 
\tau z) \eeq where $M$, $K$:  fluxes. \beq \mathcal{ G}(z)= {z \over 
2 \pi i}\ln(z) + {\rm~holomorphic}.
 \eeq
This has a supersymmetric minimum if \beq D_z W = {\partial W 
\over \partial z}+{\partial K \over \partial z}W =0 \eeq
These equations are solved 
by: \beq z \sim {\rm exp}{(-{2\pi K \over M g_s})} \eeq If the 
ratio $N/M$ is large, then $z$ is very small.  The corresponding 
space can be shown to be highly warped (in the sense of
Randall and Sundrum.\cite{Randall:1999ee})   
\beq W_o = <W>, \eeq
in this example, is exponentially small.

Including additional fluxes, it is possible to fix 
other complex structure moduli, including $\tau$.  GKP
\cite{Giddings:2001yu} provided an example with:
 \beq W= (2 
\pi)^3 \alpha^\prime[M \mathcal{G}(z)-\tau (Kz + K^\prime f(z))] 
\eeq 
$$D_\tau W = {\partial W 
\over \partial \tau} + {\partial K \over \partial \tau} W =0$$ for
$$\bar \tau = {M \mathcal{G}(0) \over K^\prime f(0)}~~~~~W_o= 2(2 \pi)^3 \alpha^\prime
M \mathcal{G}(0) $$
Note that in this example, while
$z$ is still exponentially small, and the space is highly warped, 
$W_o$ is no longer exponentially small.
It is $W_o$ which will be particularly important in what follows.

Typically, there are many additional moduli
and fluxes. 
and a possible huge number of states.
In these states, $W_o$ is essentially
a random variable.  Small $W_o$ corresponds to approximate $N=1$ 
supersymmetry, and in theses cases on can describe the
low energy physics by a supersymmetric effective 
lagrangian.  

What is the low energy physics?  In the discussion of GKP,\cite{Giddings:2001yu}
the 
radii (Kahler moduli) are not fixed. For large $R$, 
discrete shift symmetries guarantee that any dependence in W on 
the $\rho_i \sim  R^3$ is exponentially small, $e^{-c \rho}$. 
Here KKLT made a crucial observation:  exponentially small corrections,
\beq
W= W_o+ 
e^{-c\rho}
\eeq
may arise from various sources (gluino condensation, 
membrane instantons...) The resulting potential has 
supersymmetric (AdS) solutions with
\beq
D_\rho W = {\partial W \over \partial \rho} + {\partial K \over \partial
\rho}W=0~~~~~
\rho \approx -{1 \over c} \ln(W_o).
\eeq
Douglas\cite{Douglas:2003um} and Kachru\cite{kachruunpublished} gave heuristic
arguments, subsequently verified
by Douglas and Denef,\cite{Denef:2004ze}
that suggest that the distribution of $W_o$ should be essentially flat as a random
variable.  Because of the vast number of possible choices of flux, this means that there
are a vast number of states in which all of the moduli are fixed, with unbroken
supersymmetry, in a (more or less) controlled approximation.

KKLT were particularly interested
in obtaining dS spaces,
and suggested a further subset of all states would have 
supersymmetry broken: vacua with $\overline{D3}$ 
branes located at the ends of warped throats.

It is worth pausing to estimate how many states there are.
Consider a lattice of integers, (the flux lattice), with dimensionality $K$.
Denote the vectors in the lattice by  
$\vec n$.  Take $\vec n^2 \le L$ (L is the integer which appears
in the tadpole cancellation condition; that for supersymmetric
states, this translates into a condition along the lengths of vectors
can be shown, e.g., by the methods of Douglas and Denef).
If one then just evaluates the volume of a $K$ dimensional sphere of radius
$\sqrt{L}$, one expects for the number of (nearly) supersymmetric states:
$$ \mathcal{N}_{susy} \sim {L^{K/2} \over \Gamma(K/2)}.$$
Surveying known Calabi-Yau spaces, one has many examples with
$$ L \sim 1000's ~~~~~  K \sim 100's.$$

For low energy observers, physics is different in each of these 
states.  Gauge groups, coupling constants and the like
all vary. {\it The cosmological 
constant}, in particular, is a random variable in these 
$10^{1000}$(?!) states.

So, again, the
problem is not to find ``the state" which describes our universe;
this is hopeless. Instead, one needs to study statistics of these states,
and learn the distribution of 
gauge groups, matter content, couplings, cosmological constant, 
etc.

\section{Experimental Predictions from the Landscape}

With this counting, however, we have the
first striking observation:  if the landscape is correct, 
string theory can accommodate, if not explain, the small value of 
the cosmological constant.  In this setup, the cosmological constant
is essentially a random variable.  We will talk about its distribution
shortly, but if there are $10^{1000}$ states, say, there are a huge
number with cosmological constant close to that observed.

\subsection{ What data should we use (Priors)?}

Given that there is a distribution of low energy parameters, we could only hope to
predict them all from first principles if we had some cosmological principle
which would select one state.  Our working assumption is that, while cosmology is certainly
an important part of this story, all of the states of the landscape
are more or less equally likely.\footnote{More precisely,
we are assuming that if, say, states with more e-foldings of inflation are favored
over states with less,  we are actually assuming
the number of e-foldings is not highly correlated with quantities such as the number of
generations or values of low energy
couplings.  This need not be the case; one could
well imagine, for example, that supersymmetric states are more likely than generate inflation
than non-supersymmetric states, if light scalars are crucial to inflation.}
   One approach, then, would be to
take all measured parameters of Standard Model and 
cosmology as input parameters,
and ask what values of other quantities are typical, given 
these priors.
This viewpoint has been advocated by a number of authors, but
I would argue against it on several grounds.  First, consider
an analogous problem in statistical mechanics.  Suppose one had worked
out the theory of gases, and then went to examine the air in a closed
room, and discovered that at that instant, 3/4 of the atoms were in a small
corner.  We could simply accept this.  After all, this is as much an allowed state
in the ensemble as any other.  Still, it is very atypical.  We could
quantify this by asking for the expectation value of the density, and
studying fluctuations.  Similar statements apply to the landscape.
If we consider the cosmological constant, its mean value is of order
$M_p^4$ (perhaps even larger).  $\Lambda=$
is a special point; here the theory has Poincare-invariant
solutions.\footnote{A more precise analogy to
the statistical mechanical model would be provided
by asking the probability of finding universes which
admit a significant degree of complexity.}    The probability of observing a very small
cosmological constant, such as we see, is extremely tiny.  So it is necessary
to explain this fact.  Similarly, we can ask about
the values of other quantities, like $\theta_{qcd}$.
Again, in the landscape, the mean value, relative
to the CP conserving point, appears to be of order one, and the distribution
uniform.  So finding a value very close to the special point is surprising
and requires some rational explanation.

Of course, in the case of $\theta$, we can at least imagine some
microscopic explanation.  It could be that in some large subset of states
in the landscape, there are axions, and that these states are favored by
some other considerations (see below).  Perhaps approximate flavor symmetries
play some role.  For the cosmological constant, we know of nothing comparable.

\subsection{The Anthropic Principle}

Within the landscape, we have argued that there are likely to be many
states with cosmological constant similar to what we observe.  But we have
also argued that the observed value is very surprising; it is not a simple
piece of data.
This is, in fact, the most persuasive setting in which
to implement the (weak) anthropic
solution of the cosmological constant problem.\cite{Weinberg:1988cp,
Garriga:1999bf,Weinberg:2000yb}

Usually, mention of the anthropic principle brings handwringing 
about the end of science. 
But, for better or worse, the anthropic explanation is arguably 
the most plausible proposal we have to understand the small value 
of $\Lambda$.
{\it I will argue that we confront a Faustian bargain.
If we adopt the anthropic viewpoint, we are lead to the 
first predictive framework for string theory.}

This statement requires explanation on several counts.
First, I have to say that I do not know how to implement
the anthropic principle.  It is nearly impossible to say:  the weak anthropic 
principle (the requirement that we find ourselves in an 
environment or neighborhood which can support life) requires the 
cosmological constant to be..., the fine structure constant to 
be... the strength of inflationary fluctuations to be...  The problem
is simply too complicated.\cite{Aguirre:2001zx,Graesser:2004ng,Banks:2003es}

Instead, for the moment, I will adopt a more pragmatic view:
I am willing to impose,
as priors, any quantity which might plausibly be anthropic -- but 
not those which cannot be.
Examples of the former include: the gauge group, $\Lambda$, $\alpha$, 
$\Lambda_{qcd}$, $m_e, m_u,m_d$, the dark energy density.  Examples
of quantities which are 
probably not anthropic are the value of $\theta_{qcd}$ and heavy 
quark masses and mixings.

These rules may seem overly generous, 
but they leave open very real possibilities for prediction and for
falsification.
With mild (in my view) assumptions about the distribution 
of states and two anthropically motivated priors, the observed 
small cosmological constant and Higgs mass lead to a prediction 
of low energy supersymmetry.
These assumptions are true of a small piece of the landscape 
which as already been studied, but may not be true more 
generally; {\it what is important is that they can be checked}.

They also provide a possible route to more immediate falsification.
Consider $\theta_{qcd}$; it has hard
to offer any plausible anthropic explanation of its small
value. In the 
flux discretuum, on the
other hand, it appears to be a random variable with a roughly 
uniform distribution.  Some rational explanation (axions?  
$m_u=0$) is required. The mechanism must be typical of the states 
which satisfy other selection criteria in the landscape, or 
landscape idea is false.  One can speculate on possible explanations.
For example, axions might constitute the dark matter, and vacua
with axions might be selected anthropically.  But it is not clear
that even this yields a small enough $\theta$.

\section{Supersymmetry or Not}

There are some distributions which we do know, 
thanks to the work of Douglas and collaborators and Kachru
and collaborators.\cite{Denef:2004ze,Giryavets:2003vd}  Two are 
relevant to the question of low energy supersymmetry.

\begin{enumerate}
\item  $W_o$.  The distribution of $W_o$, as a complex variable, 
is known at least in some cases to be roughly uniform.  KKLT gave 
a crude argument for this, which is supported by the results of 
Douglas and Denef.  Think of
$W_o = \sum a_i n_i = 
\vec a \cdot \vec n$
where $\vec a$ is independent of the fluxes (this is the rough part of the
argument).
This gives, at small $W_o$, a uniform 
distribution of both Re $W_o$ and Im $W_o$.  So
$$\int d^2 W_o P(W_o)$$
with $P(W_o)$ approximately uniform.
\item $\tau = {1 \over g} + ia$.  Since the IIB theory has an 
$SL(2Z)$ symmetry, one might expect
$$P(\tau){d^2 \tau \over ({\rm Im}~\tau)^2}$$
with $P(\tau)$ roughly constant.  Indeed, this is what Douglas 
and Denef find. It corresponds to gauge coupling constants distributed
uniformly with $g^2$.
\end{enumerate}

\subsection{Three branches}

KKLT established that for some fraction of flux choices, one can
exhibit metastable states in a more or less systematic approximation.
In most states, no such analysis is possible, but in thinking about 
statistics, it may still sometimes be useful to use the supergravity
lagrangian and examine its solutions.
At this level of analysis,
analysis, there are three important branches of the flux landscape

\begin{enumerate}
\item  Broken supersymmetry in the leading approximation.
\item  Unbroken 
supersymmetry, $W_o \ne 0$. \item  Unbroken susy, $W_o=0$.
\end{enumerate}

Douglas and Denef suggest that the number of states on the first
branch, if one simply looks for stationary points of the
leading order potential, is infinite.  In fact, it is rather
easy to exhibit infinite sequences of gauge-invariant,
non-supersymmetric states.\cite{toappear}
Douglas and Denef argue that one should cut off the
sum over states in this case.  If one simply requires
either that the $\alpha^\prime$ corrections not be large,
or that the states be long-lived, one must cut off the sum at
small flux number ($\mathcal{O}(1)$).\cite{toappear}  So there
is no evidence from this type of counting that there are vastly
more non-supersymmetric than supersymmetric states.  On the other
hand, by their very nature, any would-be non-supersymmetric
states are difficult to explore, and it is quite possible that we
are not looking in the right place\footnote{This point has been
emphasized to me quite cogently by Silverstein}.

If it should turn
out that there are vastly more non-susy than susy states, this can overwhelm the 
usual naturalness arguments for susy.\cite{Douglas:2004qg,Susskind:2004uv}
If the non-supersymmetric branch should turn out to
dominate, this would be particularly disappointing, and not
merely for supersymmetry proponents.  As the arguments
of Douglas and Susskind make clear, there is unlikely to be any
small parameter in such a case.  Environmental selection would
simply select from a vast, complicated and essentially
inaccessible ensemble.  It is  
hard to see how any prediction should emerge.  For example,
the scenario of Arkani-Hamed and Dimopoulos\cite{Arkani-Hamed:2004fb}
can be argued not to predominate in the landscape.\cite{toappear}

Apart from arguments about the landscape, {\it nature} provides us with
reasons to hope that the picture is not so bleak.  As we have stressed,
there are many features of the Standard Model which appear neither
random nor anthropic.  Some questions must have rational explanation.

\section{The Supersymmetric, $W \ne 0$ Branch}

The supersymmetric, $W\ne 0$ branch, is distinctly more interesting.
While supersymmetry is unbroken to all orders in $\rho$, 
there is no reason to expect that this is exact.  Low energy 
dynamics are likely to break supersymmetry
in a finite number of states.\footnote{The $\overline{ D3}$ effects of KKLT
are quite possibly dual to these; in any case, as will be explained
elsewhere,\cite{toappear}
the counting of these states is similar to that which we perform now.}  
Calling $\mathcal{M}$ the scale of susy breaking ($m_{3/2} = {\mathcal{M}^2 \over 
M_p}$):
\beq
\mathcal{M}^4 = e^{-c{8\pi^2 \over g^2}}~~~~~(M_p=1)
\eeq
The uniform distribution in $g^2$ then translates into a distribtion
of $\mathcal{M}^2$ or $m_{3/2}^2$:
\beq
 P(m_{3/2}) ={dm_{3/2}^2\over 
m_{3/2}^2 (-\ln(m_{3/2}^2))}\eeq
which is roughly uniform with the logarithm of the energy 
scale.

On this branch, small cosmological constant and the facts 
just mentioned do not by themselves predict low energy supersymmetry.  We can 
ask: how many states have cosmological constant smaller than a 
give value?
As a
simplified model we write for the cosmological constant:
\beq
 \Lambda = \mathcal{M}^4 - 
3 \vert W_o \vert^2
\eeq
so that the fraction of states with cosmological constant less than $\Lambda_o$ is 
given by:
\beq
F_1(\Lambda<\Lambda_o) = \int_0^{W_{\rm 
max}} d^2 W_o  \int_{\ln(\vert W_o\vert^2)}^{\ln(\vert W_o 
\vert^2 + \Lambda_o)}
d(g^{-2})g^4
\eeq
$$~~~~~\approx \int_0^{W_{\rm max}} d^2 W_o 
{\Lambda_o \over \vert W_o \vert^2} 
(-1/\ln(W_o))^2
$$

So requiring small cosmological constant gives a
distribution of $m_{3/2}$ flat on a log scale.
But now imposing the value of the weak scale as an additional 
requirement favors supersymmetry breaking at the weak 
scale.  This is just conventional naturalness.  In terms
of the probability analysis above, the Higgs
mass will obtain contributions from many sources.  Provided
one has understood the smallness of the $\mu$ term (e.g. due
to symmetries), for a Higgs mass well below the supersymmetry
breaking scale, cancellations will be required; this corresponds
in the landscape picture to the fact that the Higgs mass
is small in a fraction of otherwise suitable states of order
$m_{higgs}^2 \over m_{susy}^2$.

\subsection{The W=0 Branch: A very low energy breaking scenario}

There is a further subset of states with $W=0$.  These will
arise in the subset of flux vacua with discrete R symmetries;
they may also arise by accident.\cite{kachruprivate}
Now one expects that both $W_o$ and $M_{susy}$ are generated 
dynamically.  For example, an exponentially small
$W$ can be generated by gluino condensation in some
other sector, which does not break susy.  So now, like $\mathcal{M}$,
the effective $W_o$ is also distributed roughly
uniformly on a log scale.   Repeating our earlier counting,
\beq
F_1 \propto \int {d^2 m_{3/2} \over m_{3/2}^4}
\eeq

So in this case, even before worrying about the value of the weak scale,
breaking of supersymmetry at the lowest possible scale is
significantly favored.  Indeed, we now have to invoke the value
of the weak scale to explain why the supersymmetry breaking
scale is not extremely small.  This is quite striking.
In past phenomenological approaches to gauge 
mediation, no particular scale for susy breaking favored by 
theoretical (naturalness) considerations.  In the landscape,
it may well turn out that the scale is necessarily quite low.
From the perspective of a model builder, this is an
example of an added input to model building

\subsection{ Possible Phenomenologies}

Each of these branches has a very different 
phenomenology.  

\begin{itemize}
\item Whether the third, possibly most
interesting branch, is favored depends on
the price of discrete symmetries in the
landscape (or $W=0$ accidentally).
Studies of this question will appear
elsewhere.\cite{kachrutoappear,toappear}
Note that, from the point of view of obtaining
the observed cosmological constant, the low energy
scenario is ``ahead" by a factor of order $10^{40}$.
Discrete symmetries may well cost more, but they may
be necessary even in the $W \ne 0$ branch to account
for proton stability and/or dark matter.

\item If discrete symmetries are costly-- the second branch, with
higher 
energy breaking, as in gravity mediation is likely. A natural 
scenario,\cite{Dine:1992yw} has  susy broken dynamically in 
a hidden sector.  Gaugino masses are then generated through anomaly 
mediation.\cite{Dine:2004is}  The hierarchy
among masses would arise because of the cosmological
moduli problem, and the need
to understand dark matter.\cite{Moroi:1999zb}\footnote{This is similar to one of the scenarios discussed by 
Arkani-Hamed and Dimopoulos.\cite{Arkani-Hamed:2004fb}}  Other scenarios, however, are possible.

\item  As we already said, if there are
overwhelmingly more non-susy than susy 
states, this is quite disappointing, for it
is very hard to see how one would make any connection 
with nature (but perhaps there are some
clues:  discrete symmetries and dark matter?  neutrino 
masses?)
\end{itemize}

\subsection{How can we settle these questions?}

What is perhaps most exciting about this story is that
we can imagine, with less than superhuman effort, answering
these questions.  We need to know:
\begin{itemize}
\item  Are there, in the leading approximation, 
exponentially more non-supersymmetric than supersymmetric vacua?  
We have indicated that the answer to this question is likely to 
be no, but we certainly cannot claim to have proven such a 
statement.  This would favor low energy supersymmetry. \item  
What is the price of discrete symmetries?  In particular, we need 
to compare the cost of suppressing proton decay and (if 
necessary) obtaining a small $\mu$ term with the price of light 
Higgs without supersymmetry ($10^{-36}$ or so), times the price 
of obtaining a stable, light dark matter particle (unknown, but 
probably not less than $10^{-36}$), times the other tunings 
required to obtain an acceptable cosmology.
\item  Is there a huge price for obtaining theories with low 
energy dynamical supersymmetry breaking?  Given the presumption 
that one can obtain a landscape of models with complicated gauge 
groups and chiral matter, it is hard to imagine that the price is 
enormous (in landscape terms).  A part in a billion, for example, 
would likely lead to a prediction of low energy supersymmetry. 
\item Are unbroken discrete {\it $R$-symmetries} at the high 
scale common?  If so, $\langle W \rangle$ must be generated 
dynamically at low energies in such vacua. In this case, we have 
seen that SUSY breaking at the lowest possible scale may be 
favored.  R symmetries might be selected by other considerations
as well, such as proton stability.

\item  
Within the present knowledge of the landscape, non-supersymmetric 
conifolds appear to be the most promising alternative to low 
energy supersymmetry.  What is the relative abundance of such 
states compared to supersymmetric states?
\end{itemize}

\section{Conclusions}

It seems likely that the landscape exists.  If so, 
at the very least, it is a very large elephant in the closet.  
What are we to make of it?  Clearly we need to explore it. {\it 
I have argued that for the first time, we may have a 
candidate predictive framework for string theory.}
\begin{itemize}
\item  The 
study of the statistics of these states has begun.  Many of the 
important questions seem accessible.\item  The proposed 
set of rules seem likely to lead to predictions.  The rules are 
subject to debate, but a sensible set of rules can probably be 
formulated. \item  Low energy supersymmetry may well be one 
output.  It is possible that we will be able to predict a 
detailed phenomenology. \item  We have about three years!
\end{itemize}

\noindent
{\bf Acknowledgements:}  I wish to thank my collaborators, Tom Banks,
Elie Gorbatov and Scott Thomas; they should not be held responsible
for some of the opinions which have been expressed here.  I also want
to thank Michael Douglas and Shamit Kachru for many conversations and
careful explanations of their results, and to thank Lenny Susskind,
Steve Weinberg and Ed Witten
for sharing their wisdom on these matters. Finally, I wish to thank
my students,
Deva O'Neil and Z. Sun.


\end{document}